\documentclass[conference, 10.1pt]{IEEEtran}
\IEEEoverridecommandlockouts
\usepackage{cite}
\usepackage{amsmath,amssymb,amsfonts}
\usepackage{algorithmic}
\usepackage{graphicx}
\usepackage{textcomp}
\usepackage{xcolor}
\usepackage{bm}
\usepackage{bbm}
\usepackage{subfigure}
\usepackage{graphicx}
\usepackage{subfig}
\usepackage{verbatim}
\usepackage{epstopdf}

\def\BibTeX{{\rm B\kern-.05em{\sc i\kern-.025em b}\kern-.08em
    T\kern-.1667em\lower.7ex\hbox{E}\kern-.125emX}}
\begin{document}

\title{Learning Proximal Operator Methods for Massive Connectivity in IoT Networks\\

}

\author{
  \IEEEauthorblockN{$\text{Yinan Zou}^{1}$, $\text{Yong Zhou}^{1}$, $\text{Yuanming Shi}^{1}$, and $\text{Xu Chen}^{2}$} 
 \IEEEauthorblockA{
 $^{1}$School of Information Science and Technology, ShanghaiTech University, Shanghai, China \\
 $^{2}$School of Computer Science and Engineering, Sun Yat-sen University, Guangzhou, China \\
  Email: \{zouyn, zhouyong, shiym\}@shanghaitech.edu.cn, chenxu35@mail.sysu.edu.cn}}

\maketitle

\begin{abstract}
Grant-free random access has the potential to support massive connectivity in Internet of Things (IoT) networks, where joint activity detection and channel estimation (JADCE) is a key issue that needs to be tackled. 
The existing methods for JADCE usually suffer from one of the following limitations: high computational complexity, ineffective in inducing sparsity, and incapable of handling complex matrix estimation. 
To mitigate all the aforementioned limitations, we in this paper develop an effective unfolding neural network framework built upon the proximal operator method to tackle the JADCE problem in IoT networks, where the base station is equipped with multiple antennas. 
Specifically, the JADCE problem is formulated as a group-sparse-matrix estimation problem, which is regularized by non-convex minimax concave penalty (MCP). 
This problem can be iteratively solved by using the proximal operator method, based on which we develop a unfolding neural network structure by parameterizing the algorithmic iterations. 
By further exploiting the coupling structure among the training parameters as well as the analytical computation, we develop two additional unfolding structures to reduce the training complexity. 
We prove that the proposed algorithm achieves a linear convergence rate. 
Results show that our proposed three unfolding structures not only achieve a faster convergence rate but also obtain a higher estimation accuracy than the baseline methods. 
\end{abstract}

\section{Introduction}
Massive machine-type communications (mMTC) is expected to provide ubiquitous wireless connectivity for billions of Internet of Things (IoT) devices \cite{al2015internet, 9163314, 9374107, 7996362}.
mMTC can support many emerging IoT applications  including smart cities and factory automation. 
mMTC has the unique features of massive connectivity and sporadic short-packet transmission. 
Applying the traditional grant-based random access in mMTC may incur excessive access delay and high signaling overhead. 
As a result, grant-free random access, as a promising candidate technology for massive connectivity, has been proposed \cite{8454392}. 
Without the need of receiving a grant from the base station (BS), each IoT device can directly send a signature sequence along with its data, thereby significantly reducing the access latency. 

Joint activity detection and channel estimation (JADCE) is recognized as an important issue for grant-free random access. 
By exploiting the device sparsity, the authors in \cite{senel2018grant} leveraged the compressed sensing (CS) approach to solve the JADCE problem, which can be regarded as a sparse signal recovery problem. 
%{\color{red}In order to apply to large-scale networks with a large number of devices, the approximate message passing (AMP) based algorithms were adopted as the main technique.
%For multiple measurement vector (MMV) problem, the authors in \cite{chen2018sparse} proposed AMP with a vector denoiser and the authors in \cite{ziniel2012efficient} proposed the AMP-MMV algorithm by using graphical model to describe the variations of the source vectors.}
Approximate message passing (AMP) algorithms were utilized to achieve sparse signal recovery for single measurement vector (SMV)  \cite{donoho2009message} and multiple measurement vector (MMV) models \cite{chen2018sparse}. 
While the AMP-based approaches are effective at recovering sparse signal, they may not be able to converge when the signature sequence is either non-Gaussian or mildly ill-conditioned \cite{rangan2019convergence}. 
The authors in \cite{liang2021sparse} regarded the JADCE problem as a mixed integer programming (MIP) problem and achieve an optimal solution by leveraging branch-and-bound approach.
Moreover, the authors in \cite{jiang2018joint} formulated the JADCE problem as a group LASSO problem, and adopted the iterative shrinkage thresholding algorithm (ISTA) to solve it \cite{qin2013efficient}. 
Nevertheless, ISTA usually requires hundreds of iterations to converge \cite{beck2009fast}, and thus is not suitable for low-latency applications, especially when there are a lot of 
IoT devices. 

Many efforts have recently been devoted to develop deep learning based methods to recover sparse signal for SMV models. 
A common strategy is to unfold the state-of-the-art iterative algorithms as recurrent  neural networks (RNN)  \cite{gregor2010learning, chen2018theoretical, liu2019alista}. 
Specifically, the authors in \cite{gregor2010learning} proposed a learned ISTA (LISTA) framework, which trains a neural network to learn the threshold parameters and the weight matrices of ISTA. 
The linear convergence rate of LISTA was further proved in \cite{chen2018theoretical}. 
The authors in \cite{liu2019alista} derived an analytic LISTA model, which simplifies the training process by only learning the step-size and threshold parameters. 
The authors in \cite{shi2020sparse}, \cite{9508782} developed an efficient JADCE algorithm, which adopts the mixed $\ell_1/\ell_2$-norm as regularization. 
However, the above studies solved the sparse signal recovery problem regularized by either $\ell_1$-norm or $\ell_1/\ell_2$-norm, which is less capable of inducing sparsity than the non-convex sparsity-inducing penalties (SIPs). 
Thus, the authors in \cite{yang2020learning} proposed a learning proximal operator method (LePOM), where non-convex SIPs were adopted for regularization. 
However, this method is only effective in sparse real signal recovery and cannot be directly extended to tackle MMV models.
In summary, the aforementioned methods suffer from one of the following limitations: high computational complexity, ineffective in inducing sparsity, and incapable of handling complex matrix estimation.

% \textcolor{blue}{Nonetheless, LePOM was developed based on scalar shrinkage-thresholding operator (SSTO), and thus cannot be directly applied to solve the complex group-sparse-matrix estimation problem in IoT networks.}  

%{\color{red}In order to simplify training, \cite{liu2019alista} proposed Analytic LISTA (ALISTA) where weight matrices can be obtained by solving a optimization problem. Thus training parameters reduce to the stepsize and thresholding parameters. 
%The $\ell_0$ norm can directly measure the sparsity, and the $\ell_1$ norm is the convex relaxation of the $\ell_0$ norm. Since the $\ell_0$ norm regularized optimization problem is hard to handle, lots of works usually utilize the $\ell_1$ norm  to induce sparsity. 
%Nonconvex SIPs are a better choice to be a penalty since they are better approximation of the $\ell_0$ norm compared with the $\ell_1$ norm \cite{eldar2012compressed}.
%In order to promote sparseness more effectively, learning proximal operator method (LePOM) was proposed by utilizing nonconvex SIP which is more favorable in inducing sparsity than the $\ell_1$ norm \cite{yang2020learning}. However, the above methods cannot be applied for JADCE problem because SSTO is not directly for group-sparse matrix. Therefore, the authors in \cite{shi2020sparse} developed LISTA for group-sparse matrix estimation problem (LISTA-GS) by adopting multidimensional shrinkage thresholding operator (MSTO) \cite{puig2011multidimensional}.}

In this paper, we carry out research into the JADCE problem in a single-cell IoT network, where grant-free random access is adopted and many IoT devices transmit non-orthogonal but unique signature sequences to the BS in a sporadic manner. 
We formulate the JADCE problem as a complex group-sparse matrix estimation problem, which cannot be directly tackled by the state-of-the-art methods developed based on scalar shrinkage-thresholding operator (SSTO) (e.g., LISTA \cite{gregor2010learning} and LePOM \cite{yang2020learning}). 
To enable faster convergence, we adopt the minimax concave penalty (MCP) regularization, which is more capable of promoting sparsity than the widely-used $\ell_1$-norm regularization. 
%Due to the sum form of SIP, MCP can be extended to solve sparse matrix recovery problem with arbitrary sparse structures, and is not limited to group-row-sparsity case. 
In addition, we develop an efficient unfolding neural network structure, termed learned proximal operator method for complex group-row-sparsity (LPOM-GS), which adopts the MCP regularization to tackle the complex group-sparse matrix estimation problem. 
To reduce the training complexity, we further develop two variants of LPOM-GS by exploiting the coupling structure among the trainable parameters as well as the analytical computation. 
Moreover, we theoretically prove that our proposed algorithm can achieve a linear convergence rate. 
Simulations demonstrate the superior performance of the proposed three unfolding structures in terms of the estimation accuracy and the convergence rate. 
Results also show the robustness of the proposed algorithm under different signature sequences.

%Compared with recovering the sparse vector, recover group-row-sparse matrix breaks the independency between the sparse vectors, which brings challenge to use an appropriate sparseness measure to characterize the group-sparse matrix. We address this challenge by introducing a nonconvex SIP for group-sparse matrix. 
%In order to achieve a faster convergence rate and promote more sparseness, we propose a new unrolled deep neural network named learned proximal operator method for group-row-sparsity (LPOM-GS). Inspired by LISTA, the proposed algorithm unfolds proximal operator method corresponding to SIP-regularized problems. We use the group-sparse inducing regularizer and formulate a group-sparse matrix estimation problem. Inspired by partial weight coupling phenomenon, we propose LPOM-GSCP to reduce trainable parameters. An improved network structure named Analytic LPOM-GS (ALPOM-GS) is further proposed that weight matrix can be obtained by solving an optimization problem. 
%Simulations demonstrate the superior performance of our algorithm. 

%This rest of the paper is organized as follows.
%In Section II, we describe the system model and problem formulation. 
%We present the proposed unfolding framework based on the proximal operator with three different network structures. %Simulation results of the proposed method are provided in Section IV. Finally, Section V concludes this paper.
 
\section{System Model and Problem Formulation}
\subsection{System Model}
Consider a single-cell IoT network, where one $ M $-antenna BS serves $ N $ single-antenna IoT devices using the grant-free uplink transmission. 
Without loss of generality, we assume that the number of IoT devices (i.e., $N$) is much greater than the number of antennas (i.e., $M$) deployed at the BS. 
As the IoT devices transmit sporadically, we assume that each IoT device independently decides whether or not to transmit in each transmission block. 
In each transmission block, we denote $a_n = 1$, if device $n$ is active, and $a_n = 0$ otherwise. 
As grant-free uplink transmission is considered, an active device, without the need of obtaining a scheduling grant, directly transmits its signature sequence in conjunction with its data to the BS, while an inactive IoT device keeps silent. 
We denote $ s_n(\ell) $ as the $ \ell $-th signature symbol transmitted by device $ n $ and $ \bm {h}_n \in  \mathbb{C}^M$ as the channel coefficient vector between the BS and IoT device $ n $. 
By assuming symbol-level synchronized transmission from the active IoT devices, the signature sequence superimposed at the BS, denoted by $ \bm{y}(\ell) \in \mathbb{C}^M $, is given by 
\begin{equation}\label{key1}
\bm{y}(\ell) = \sum_{n=1}^{N} \bm{h}_n a_n s_n(\ell) + \bm{z}(\ell),\quad  \ell = 1,\ldots, L,
\end{equation}
where $L$ is the length of the signature sequence and $ \bm{z}(\ell) \in \mathbb{C}^M $ is the additive white Gaussian noise (AWGN) with zero mean and variance $ \sigma^2$.
We follow \cite{chen2018sparse} and consider quasi-static block fading, where the channel coefficient vector of each link (e.g., $\bm h_n$) remains invariant within one transmission block but varies independently across different transmission blocks. 

To ensure low-overhead grant-free uplink transmission,  the length of the signature sequence is generally much smaller than the number of IoT devices. 
Hence, we cannot assign orthogonal signature sequences to all IoT devices. 
We follow \cite{shi2020sparse} and assume that  a non-orthogonal but unique signature sequence (e.g., $\bm s_n \in  \mathbb{C}^L$) is generated for each IoT device according to a certain distribution. 
In the simulations, we consider two different preamble signatures, i.e., complex Gaussian matrix and binary matrix, to demonstrate the robustness of the proposed method. 
We assume that the BS knows these unique signature sequences. 

For ease of notations, we rewrite the signature sequence  superimposed at the BS in the matrix form as 
\begin{equation} \label{key2}
\bm{Y} = \bm{SAH} + \bm{Z},
\end{equation}
where $ \bm{Y} = [\bm{y}(1), \ldots, \bm{y}(L)]^{\mathrm{T}} \in \mathbb{C}^{L \times M}$, $ \bm{S}  =[\bm{s}(1), \ldots, \bm{s}(L)]^{\mathrm{T}} \in \mathbb{C}^{L \times N} $ with $ \bm{s}(\ell) = [s_1(\ell), \ldots, s_N(\ell)]^{\mathrm{T}} \in \mathbb{C}^N$, $ \bm A = \operatorname{Diag}(a_1, \ldots, a_N) \in \mathbb{R} ^{N\times N} $, $ \bm{H}  = [ \bm{h}_1, \ldots, \bm{h}_N]^{\mathrm{T}} \in \mathbb{C}^{N\times M}$,  and $ \bm{Z}  =[\bm{z}(1), \ldots, \bm{z}(L)]^{\mathrm{T}} \in \mathbb{C}^{L \times M} $. 
Before decoding data from each active IoT device, we need to solve the JADCE problem, i.e., detecting the device activity matrix $\bm A$ and estimating the channel coefficient matrix $\bm H$. 
By denoting $ \bm{X} = \bm{AH} \in \mathbb{C}^{N\times M} $, we can rewrite (\ref{key2}) as 
\begin{equation}\label{key3}
\bm{Y}  =\bm{SX} + \bm{Z}. 
\end{equation}

%For ease of  notation, we denote $ \mathbf{Y} = [\mathbf{y}(1), \cdots, \mathbf{y}(L)]^T \in \mathbb{C}^{L \times M}$ as the received pilot signal matrix across $ M $ antennas over the transmission duration of $ L $ symbols, $ \mathbf{H}  = [ \mathbf{h}_1, \cdots, \mathbf{h}_N]^T \in \mathbb{C}^{N\times M}$ as the channel matrix between $ N  $ devices and $ M $ BS antennas, $ \mathbf{Z}  =[\mathbf{z}(1), \cdots, \mathbf{z}(L)] \in \mathbb{C}^{L \times M} $  as the noise matrix at the BS, and $ \mathbf{S}  =[\mathbf{s}(1), \cdots, \mathbf{s}(L)] \in \mathbb{C}^{L \times N} $ as the known signature matrix with $ \mathbf{s}(\ell) = [s_1(\ell), \cdots, s_N(\ell)]^T \in \mathbb{C}^N $. As a result, the pilot signal received at the BS can be rewritten as
%\begin{equation} \label{key2}
%\mathbf{Y} = \mathbf{SAH} + \mathbf{Z},
%\end{equation}

\subsection{Problem Formulation}
As $\bm A$ is a diagonal matrix, $\bm{X} $ has the structure of group-sparsity in rows. 
This implies that each column of matrix $\bm{X}$ shares the same support. 
The matrix estimation problem can be formulated as \cite{yuan2006model}
\begin{equation} \label{key4}
\begin{aligned}
\mathcal{P} : \mathop{ \text{minimize}}_{\bm{X}\in\mathbb{C}^{N \times M} }
\frac{1}{2} \left\| \bm{Y-SX} \right\|_F^2 + \lambda G(\bm{X}),
\end{aligned}
\end{equation}
where $ \lambda > 0 $ denotes the regularization parameter. 
Note that $ G(\mathbf{X})  $ is a sparsity-inducing penalty that is added to promote the group-row-sparsity of matrix $\bm X$. 
We rewrite (\ref{key3}) as its real-valued counterpart as follows
\begin{equation} \label{key6}
\begin{aligned}
\tilde{\bm{Y}}&=\tilde{\bm{S}}\tilde{\bm{X}}+\tilde{\bm{Z}}=
\begin{bmatrix} \mathcal{R}\left\{\bm{S}\right\} & -\mathcal{I}\left\{\bm{S}\right\} \\ \mathcal{I}\left\{\bm{S}\right\} & \mathcal{R}\left\{\bm{S}\right\} \end{bmatrix}
\begin{bmatrix} \mathcal{R}\left\{\bm{X}\right\}  \\  \mathcal{I}\left\{\bm{X}\right\} \end{bmatrix}
+\begin{bmatrix} \mathcal{R}\left\{\bm{Z}\right\}  \\  \mathcal{I}\left\{\bm{Z}\right\} \end{bmatrix},
\end{aligned}
\end{equation}
where $\mathcal{R}\{\cdot\}$ and $\mathcal{I}\{\cdot\}$ denote the real and imaginary parts of a complex matrix. 
Based on (\ref{key6}), we can rewrite problem $\mathcal{P}$ as
\begin{equation} \label{key7}
\begin{aligned}
\mathcal{P}_r:\mathop{\text{minimize}}_{\tilde{\bm{X}}\in\mathbb{R}^{2N \times M}}
\frac{1}{2}\|\tilde{\bm{Y}}-\tilde{\bm{S}}\tilde{\bm{X}}\|_F^2+\lambda G(\tilde{\bm{X}}).
\end{aligned}
\end{equation}

Although ISTA for group-row-sparsity (ISTA-GS) \cite{yuan2006model} can be utilized to tackle problem (\ref{key7}), it suffers from the following two limitations. 
First, ISTA-GS usually requires hundreds of iterations to converge due to its sublinear convergence rate. 
Second, the estimation performance of ISTA-GS may be severely degraded if the regularization parameter $\lambda$ is not appropriately selected. 
In addition, the existing LISTA \cite{gregor2010learning} and LePOM \cite{yang2020learning} methods were developed based on SSTO, and hence cannot be directly utilized to tackle the JADCE problem for complex matrix estimation. 
While utilizing LISTA-GS \cite{shi2020sparse} can solve the complex matrix estimation problem, LISTA-GS is only capable of solving $\ell_1 / \ell_2$-norm regularized problems, which is not as effective as many non-convex SIPs in promoting sparsity.

\section{Proposed Unfolding Structures}
In this section, we present three neural network structures that unfold the proximal operator method and tackle the matrix estimation problem regularized by MCP. 

\subsection{Unfolding Structure I: LPOM-GS}
For (\ref{key7}), we perform the following operation iteratively to recover the group-row-sparse matrix $\tilde{\bm{X}}$\cite{yuan2006model}
\begin{equation} \label{key8}
\begin{aligned}
\tilde{\bm{X}}^{k+1}=P_{\lambda \gamma_k}\left(\tilde{\bm{X}}^k + \gamma_k\tilde{\bm{S}}^T(\tilde{\bm{Y}} - \tilde{\bm{S}} \tilde{\bm{X}}^k   ) \right),
\end{aligned}
\end{equation}
where $\tilde{\bm{X}}^{k}$ denotes the estimate of matrix $\tilde{\bm{X}}$ at the $k$-th iteration, $\gamma_k$ denotes the step size, and $P_{\lambda \gamma_k}(\cdot)$ denotes the proximal operator for sparsity inducing.
By denoting $\bm{W}^k=\bm{I}-\gamma_k\tilde{\bm{S}}^T\tilde{\bm{S}}$ and $\bm{B}^k=\gamma_k\tilde{\bm{S}}^T$, we rewrite (\ref{key8}) as
\begin{equation}\label{key9}
\begin{aligned}
\tilde{\bm{X}}^{k+1}=P_{\lambda \gamma_k}\left(\bm{W}^k\tilde{\bm{X}}^k + \bm{B}^k\tilde{\bm{Y}}  \right).
\end{aligned}
\end{equation}

Note that we can view (\ref{key9}) as a one-layer neural network by regarding $\tilde{\bm{X}}^k$ as the input and $\tilde{\bm{X}}^{k+1}$ as the output, $P_{\lambda \gamma_k}$ as the activation function, and $\bm{W}^k$ and $\bm{B}^k$ as the weight parameters. 
We model the $K$ iterations of (\ref{key9}) as a $K$-layer RNN via cascading the neural network layer by layer. 
Based on the proximal operator method, we develop the following unfolding neural network to recover the group-row-sparse matrix
\begin{equation} \label{key10}
\begin{aligned}
\tilde{\bm{X}}^{k+1}=P_{\theta_k,G_{\eta_k}}\left(\bm{W}^k\tilde{\bm{X}}^k + \bm{B}^k\tilde{\bm{Y}}  \right), \quad k = 0,\ldots,K-1,
\end{aligned}
\end{equation}
where $\theta_k = \lambda \gamma_k$ denotes the thresholding parameter at the $k$-th layer, $\eta_k$ denotes the non-convexity measure, $G_{\eta_k}(\cdot)$ denotes a specific SIP, and $P_{\theta,G_{\eta_k}}(\cdot)$ denotes the following activation function \cite{yang2020learning}
\begin{equation} \label{key11}
\begin{aligned}
P_{\theta_k,G_{\eta_k}}(\bm{U})=\text{arg}\mathop{\text{min}}_{\bm{X}}\theta_k G_{\eta_k}(\bm{X})+\frac{1}{2}\left\| \bm{X}-\bm{U} \right\|^2_F.
\end{aligned}
\end{equation}

%\begin{equation}\label{key5}
%\begin{aligned}
%F(\mathbf{X})=\sum_{i=1}^N \sum_{j=1}^{M} f(\mathbf{X}_{i,j}).
%\end{aligned}
%\end{equation}

By defining $G_{\eta_k}(\bm{Z}) = \sum_{i,j}g_{\eta_k}(\bm{Z}_{i,j}) $ where $g_{\eta_k}(\cdot)$ is a weakly convex sparseness measure, problem (\ref{key11}) can be decomposed into the following multiple independent univariate optimization problems
\begin{equation}\label{key12}
\begin{aligned}
\left(P_{\theta_k,G_{\eta_k}}(\bm{U})\right)_{i,j}&=\text{arg}\mathop{\text{min}}_{\bm{X}_{i,j}}\theta_k g_{\eta_k}(\bm{X}_{i,j}) +\frac{1}{2}\left( \bm{X}_{i,j}-\bm{U}_{i,j} \right)^2.
\end{aligned}
\end{equation}

\begin{figure}[tbp]
	\centering
	\includegraphics[width=0.45\textwidth]{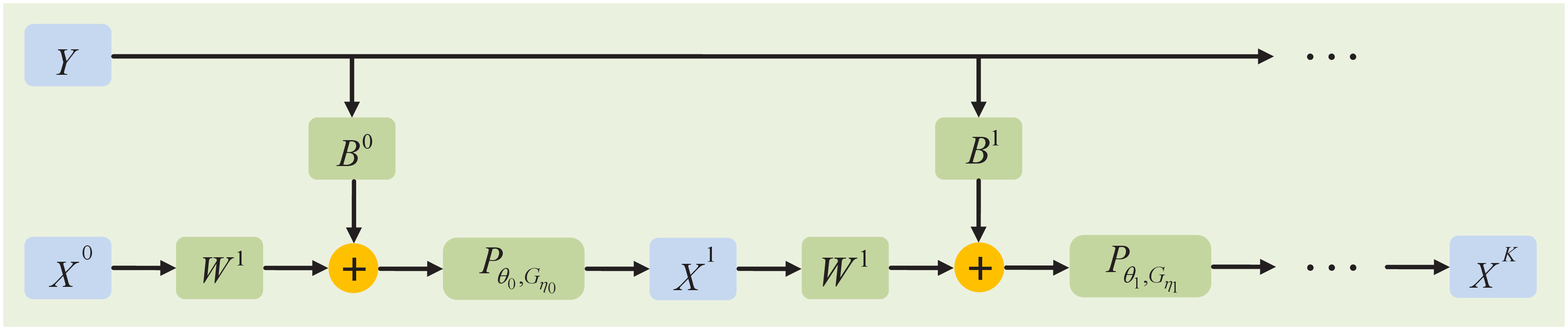}
	\captionsetup{font={footnotesize}}
	\caption{Illustration of proposed LPOM-GS structure with trainable parameters $ \{ \bm{W}^k, \bm{B}^k, \theta_k, \eta_k \}_{k=0}^{K-1}$ .} 
	\label{fig1}
	\vspace{-0.5cm}
\end{figure}

Problem (\ref{key12}) is a proximal operator problem, where different SIPs can be adopted to promote sparsity. %such as $\ell_1$-norm, MCP, and smoothly clipped absolute deviation (SCAD). 
As $\ell_1$-norm is less effective in promoting sparsity than MCP, and smoothly clipped absolute deviation is generally much more difficult to be trained than MCP, we choose MCP as the SIP in this paper \cite{zhang2010nearly}. 
Based on the definition of MCP, we have 
\begin{equation}
	g_{\eta_k}(\bm{X}_{i,j}) = \left \{
	\begin{aligned}	
		& | \bm{X}_{i,j} | - \eta_k \bm{X}_{i,j}^2,&| \bm{X}_{i,j} | \leq \frac{1}{2\eta_k}, \\
		& \frac{1}{4\eta_k}, & | \bm{X}_{i,j} | > \frac{1}{2\eta_k}. 
	\end{aligned}
	\right. \label{key13}
\end{equation}

Note that $g_{\eta_k}(\cdot)$ is even, differentiable and satisfies $\text{lim}_{z \rightarrow 0^+} \frac{g_{\eta_k}(z)}{z} = \alpha < +\infty $. Besides, $ g_{\eta_k}(\cdot) $ and $\frac{g_{\eta_k}(z)}{z}$ are non-decreasing and non-increasing functions over $[0,+\infty)$ and $(0,+\infty)$, respectively. 
As a result,  (\ref{key12}) can be expressed as 
\begin{equation}
	\label{proximal}
	\begin{aligned}
		&\left(P_{\theta_k,G_{\eta_k}}(\bm{U})\right)_{i,j} = \text{arg}\mathop{\text{min}}_{\bm{X}_{i,j}}
		\theta_k \bigg( (| \bm{X}_{i,j} |-\eta_k \bm{X}_{i,j}^2)  \mathbbm{1}_{| \bm{X}_{i,j} |\leq \frac{1}{2\eta}} \\ & (\bm{X}_{i,j}) 
		+\frac{1}{4\eta_k}\mathbbm{1}_{| \bm{X}_{i,j} | > \frac{1}{2\eta_k}} (\bm{X}_{i,j}) \bigg) +\frac{1}{2}\left( \bm{X}_{i,j}-\bm{U}_{i,j} \right)^2. 
	\end{aligned}
\end{equation} 

%\begin{equation}
%\begin{aligned}
%f(\{X_{i,j}})= (\lvert \mathbf{X}_{i,j} \rvert - \eta \mathbf{X}_{i,j}^2)\mathbbm{1}_{\lvert \mathbf{X}_{i,j} \rvert \leq \frac{1}{2\eta}}(\mathbf{X}_{i,j}) \\
%+\frac{1}{4\eta}\mathbbm{1}_{\lvert \mathbf{X}_{i,j} \rvert > \frac{1}{2\eta}}(\mathbf{X}_{i,j}) \nonumber
%\end{aligned} 
%\end{equation}

To ensure that $( P_{\theta_k,G_{\eta_k}}(\bm{U}) )_{i,j}$ is continuously differentiable, $\eta_k $ is required to be smaller than $ \frac{1}{2\theta_k}$. 
For notational simplicity, we denote $z(\bm{X}_{i,j}) = \frac{(\bm{X}_{i,j}-\bm{U}_{i,j})^2}{2}$ and $(P_{\theta_k,G_{\eta_k}}(\bm{U}))_{i,j} = \text{arg}\mathop{\text{min}}_{\bm{X}_{i,j}} h(\bm{X}_{i,j})$, where $h(\bm{X}_{i,j})$ is 
\begin{equation}
	h(\bm{X}_{i,j}) =  
	 \left \{
	\begin{aligned}
		& z(\bm{X}_{i,j})\! +\! \theta_k(\bm{X}_{i,j}-\eta_k \bm{X}_{i,j}^2),    \; 0\leq \bm{X}_{i,j} \leq \frac{1}{2\eta_k}, \\
		& z(\bm{X}_{i,j})\! -\! \theta_k(\bm{X}_{i,j}+\eta_k \bm{X}_{i,j}^2),   \frac{-1}{2\eta_k} \leq \bm{X}_{i,j}  <0, \\
		& z(\bm{X}_{i,j}) +\frac{\theta_k}{4\eta_k}, \qquad  \qquad \qquad \text{otherwise}.
	\end{aligned}
	\right. \label{key14}
\end{equation}
By setting $h^{'}(\bm{X}_{i,j}) = 0$, we have $\bm{X}_{i,j} = \frac{\bm{U}_{i,j}-\theta_k}{1-2\theta_k\eta_k}$ if $\bm{X}_{i,j} \in [0, \frac{1}{2\eta_k}]$, $\bm{X}_{i,j} = \frac{\bm{U}_{i,j}+\theta_k}{1-2\theta_k\eta_k}$ if $ \bm{X}_{i,j} \in [-\frac{1}{2\eta_k}, 0)$, and $ \bm{X}_{i,j} = \bm{U}_{i,j}$ otherwise. 
Combining all these cases, we have $\bm{X}_{i,j} \in \{0,\frac{\bm{U}_{i,j}-\theta_k}{1-2\theta_k\eta_k},\frac{\bm{U}_{i,j}+\theta_k}{1-2\theta_k\eta_k},\bm{U}_{i,j}\}. $  After some simple mathematical manipulations, we obtain the following activation function 
%\begin{equation}
%\begin{aligned}
%\left( P_{\theta}(\mathbf{U}) \right)_{i,j}=\frac{\mathbf{U}_{i,j}-\theta \text{sign}(\mathbf{U}_{i,j})}{1-2\theta \eta}\mathbbm{1}_{\theta < | \mathbf{U}_{i,j} | \leq %\frac{1}{2\eta}}(\mathbf{U}_{i,j}) \\
%+\mathbf{U}_{i,j}\mathbbm{1}_{| \mathbf{U}_{i,j} | > \frac{1}{2\eta}}(\mathbf{U}_{i,j}) 
%\end{aligned}
%\end{equation}
\begin{equation}
	( P_{\theta_k,G_{\eta_k}}(\bm{U}) )_{i,j} = \left \{
	\begin{aligned}
		& 0,  
		\qquad  \qquad \qquad   \quad \; |\bm{U}_{i,j}| \leq \theta_k , \\
		& \frac{\bm{U}_{i,j}\!-\!\theta_k \text{sign}(\bm{U}_{i,j})}{1-2\theta_k \eta_k}, \theta_k \! <\! | \bm{U}_{i,j} |\! \leq\! \frac{1}{2\eta_k}, \\
		& \bm{U}_{i,j},
		\quad \qquad \qquad  \quad       | \bm{U}_{i,j} | > \frac{1}{2\eta_k}.
	\end{aligned}
	\right. \label{key14}
\end{equation}
By substituting (\ref{key14}) into (\ref{key10}), we obtain the unfolding neural network structure in Fig. \ref{fig1}, termed LPOM-GS. 
The trainable parameters are denoted as $ \bm{\Theta} = \{ \bm{W}^k$, $\bm{B}^k$, $\theta_k, \eta_k \}_{k=0}^{K-1}$. 

\begin{figure}[tbp] 
	\centering
	\includegraphics[width=0.4\textwidth]{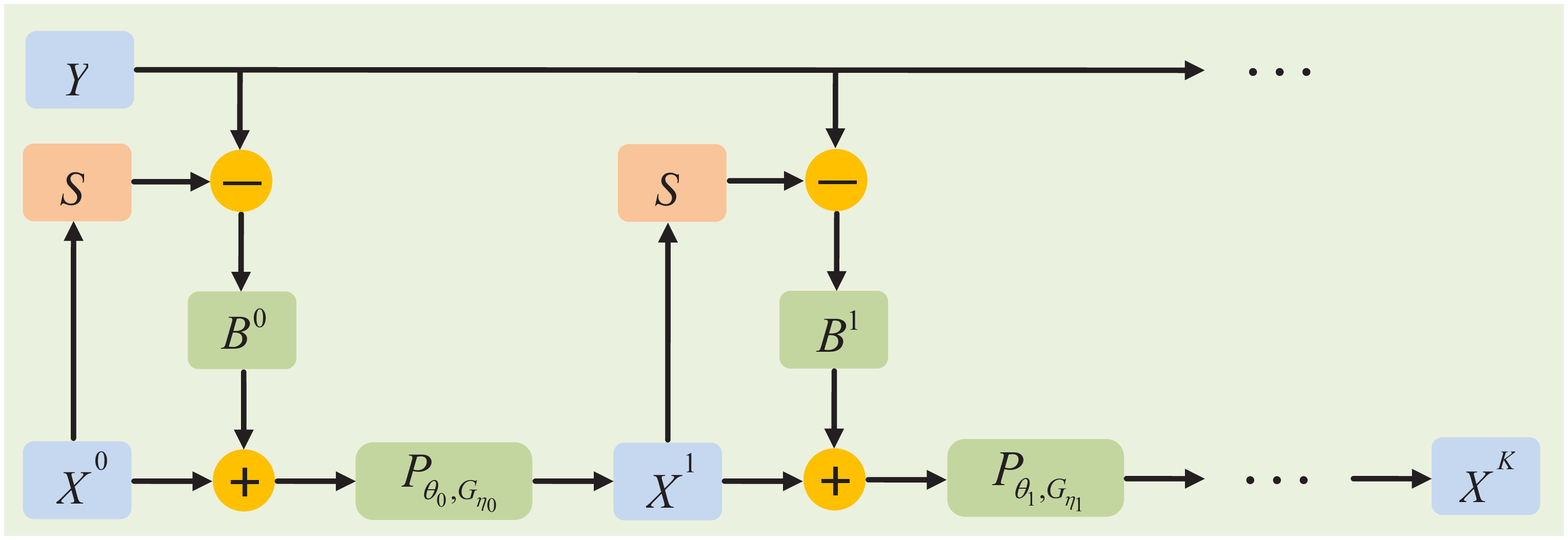}
	\captionsetup{font={footnotesize}}
	\caption{Illustration of proposed LPOMCP-GS structure with trainable parameters $ \{ \bm{B}^k, \theta_k, \eta_k \}_{k=0}^{K-1}$.} 
	\label{LPOM-GSCP}
	\vspace{-0.5cm}
\end{figure}

%When the number of layers is $K$, the total number of trainable parameters is $K(4N^2+4NL+2)$. 

%The key idea of LPOM-GS is to view $ \bm{\Theta} = \{ \bm{W}^k$, $\bm{B}^k$, $\theta_k, \eta_k \}_{k=0}^{K-1}$ as trainable parameters and the number of trainalbe parameters is $K(4N^2+4NL+2)$. Fig. \ref{fig1} shows the structure of the proposed LPOM-GS.

\subsection{Unfolding Structure II: LPOMCP-GS}
As the number of trainable parameters of LPOM-GS can be large, an overfitting issue may occur. 
%Inspired by \cite{chen2018theoretical}, 
We exploit the coupling relationship between weight matrices $\bm{W}^k$ and $\bm{B}^k$ to reduce the number of trainable parameters, thereby reducing the probability of overfitting. 
According to \cite[Theorem 1]{chen2018theoretical}, $\bm{W}^k$ and $\bm{B}^k$ asymptotically satisfy the following coupling structure
\begin{equation} \label{key15}
	\begin{aligned}
 \bm{W}^k = \bm{I} - \bm{B}^k \tilde{\bm{S}}.
	\end{aligned}
\end{equation}
We apply the coupling structure (\ref{key15}) into LPOM-GS to reduce the number of trainable parameters. 
The simplified neural network structure, termed LPOMCP-GS, is given by
\begin{equation} \label{eqn15}
	\begin{aligned}
		\tilde{\bm{X}}^{k+1}=P_{\theta_k,G_{\eta_k}}\left(\tilde{\bm{X}}^k\! +\! \bm{B}^k(  \tilde{\bm{Y}} - \tilde{\bm{S}} \tilde{\bm{X}}^k ) \right)\!,\! k = \! 0,\ldots,K-1,
	\end{aligned}
\end{equation}
where $ \bm{\Theta} = \{  \bm{B}^k, \theta_k , \eta_k \}_{k=0}^{K-1} $ are trainable parameters. 
We plot the structure of the proposed LPOMCP-GS in Fig. \ref{LPOM-GSCP}.
%For LPOM-GSCP with $K$ layers, the number of trainable parameters is $K(4NL+2)$. 

In the following, we prove that $\tilde{\bm{X}}^k$ has the no-false-positive property and LPOMCP-GS can achieve a linear convergence rate. 
We denote $\psi(\tilde{\bm{X}})=[ \|\tilde{\bm{X}}_{1,:} \|_2 ,\ldots,\|\tilde{\bm{X}}_{2N,:} \|_2  ]^T$.
By denoting $ \tilde{\bm{X}}^{*}$ as the ground truth, we define 
$\mathcal{X}(\mu_x,s,\epsilon)=\{ (\tilde{\bm{X}}^*,\tilde{\bm{Z}})| \| \tilde{ \bm{X}}^*_{i,:} \|_{2} \leq \mu_x, \forall i, \text{supp}(\psi(\tilde{\bm{X}}^{*}))\leq s, \| \tilde{\bm{Z}} \|_F \leq \epsilon  \}$. 

\textbf{Theorem 1} 
We set the input of LPOMCP-GS as $\tilde{\bm{Y}}=\tilde{\bm{S}}\tilde{\bm{X}}^*+\tilde{\bm{Z}}$ and $\tilde{\bm{X}}^0=\bm{0}$.
We denote $\{\tilde{\bm{X}}^k\}_{k=1}^{\infty}$ 
as the output of LPOMCP-GS,
$\text{supp}(\psi(\tilde{\bm{X}}^{*}))$ as $S$ and $\|\bm{X}\|_{2,1}=\sum_n\|\bm{X}_{n,:}\|_2$. 
If $\|\bm{B}^k\|_{2,1}\leq \mu_{B}$, $\| \tilde{ \bm{X}}^*_{i,:} \|_2 \leq \mu_x $, $\forall i$, $\| \tilde{\bm{Z}} \|_F \leq \epsilon$ , $|S| \leq s$  and 
\begin{equation}\label{proof1}
	\begin{aligned}
		 \frac{1}{\alpha} \left(  \phi \mathop{\text{sup}}_{(\tilde{\bm{X}^*},\tilde{\bm{Z}})\in\mathcal{X}(\mu_x,s,\epsilon)  }  \| \tilde{\bm{X}}^k  - \tilde{ \bm{X}}^* \|_{2,1} + \mu_{B} \epsilon \right) = \theta_k \leq \frac{1}{2 \eta_k},
	\end{aligned}
\end{equation}
\begin{equation}
	\begin{aligned}
		\bm{B}^k \in  \mathop{ \text{argmin}}_{\bm{B} \in \mathbb{R}^{2N\times 2L} }  \{ & \max \| \bm{B}\|_2 | \bm{B}_{i,:}\tilde{\bm{S}}_{:,i}=1, \forall i,  \\ & \mathop{\max}_{i \neq j} | \bm{B}_{i,:}\tilde{\bm{S}}_{:,j} | = \phi \},
	\end{aligned}
\end{equation}
where
$
	\phi = \inf_{ \substack{ \bm{B} \in \mathbb{R}^{2N\times 2L} \\ \bm{B}_{i,:}\tilde{\bm{S}}_{:,i}=1, \forall i  } }  \mathop{\max}_{i \neq j } | \bm{B}[i,:]\tilde{\bm{S}}[:,j] |,
$
then
\begin{equation}
	\text{supp}(\psi (\tilde{ \bm{X}}^{k})) \subseteq \text{supp}(\psi (\tilde{\bm{X}}^{*})) .
\end{equation}

%
%\textbf{Definition 2}
%If $G_\eta(\cdot): \mathbb{R}^{N\times M}\rightarrow \mathbb{R}_+$ is locally Lipschitz, then its generalized gradient is given by:
%\begin{equation}
%	\begin{aligned}
%	& \partial G_\eta(\bm{Z}) = \{G_\eta^{'}(\bm{Z})| \langle G_\eta^{'}(\bm{Z}),\bm{X} \rangle \leq \\
%	& \mathop{\text{lim}}_{\theta\to0^+} \frac{G_\eta(\bm{Z}+\theta\bm{X})-G_\eta(\bm{Z})}{\theta}, \forall \bm{X} \in \mathbb{R}^{N\times M} \}
%\end{aligned}
%\end{equation}
%
%\textbf{Lemma 1}
%For weakly convex SIP $G_{\eta}(\bm{Z}) = \sum_{i,j}g_{\eta}(\bm{Z}_{i,j}) $,  there exists $G^{'}_\eta(\bm{Z}) \in \partial G_\eta(\bm{Z}) $, such that $\| G^{'}_\eta(\bm{Z}) \|_F \leq \sqrt{\omega}\alpha$, where $\omega$ denotes the number of the elements of $\bm{Z}$.
%
%\emph{Proof}: See Appendix B. \hfill $\blacksquare$ 

\textbf{Theorem 2}  
We set the input of LPOMCP-GS as $\tilde{\bm{Y}}=\tilde{\bm{S}}\tilde{\bm{X}}^*+\tilde{\bm{Z}}$ and $\tilde{\bm{X}}^0=\bm{0}$.
We denote $\{\tilde{\bm{X}}^k\}_{k=1}^{\infty}$ 
as the output of LPOMCP-GS, and
 $\text{supp}(\psi(\tilde{\bm{X}}^{*}))$ as $S$. 
 For $\| \tilde{ \bm{X}}^*_{i,:} \|_{2} \leq \mu_x, \forall i, \| \tilde{\bm{Z}} \|_F \leq \epsilon $, and $S\leq s$, we obtain 
\begin{equation}
	\begin{aligned}
		\| \tilde{\bm{X}}^{k} - \tilde{\bm{X}}^{*} \|_F \leq s \mu_x \text{exp}(-c_1 k) + \epsilon c_2. 
	\end{aligned}
\end{equation}
where $c_1>0$ and $c_2>0$.

Theorem 1 indicates that the index set of the rows containing non-zero elements belongs to the support of $\psi (\tilde{\bm{X}}^{*})$. 
Theorem 2 demonstrates that LPOMCP-GS can recover the complex sparse matrix with a linear convergence rate $\mathcal{O} (\text{log}( \frac{1}{\epsilon} )) $  in a noisy scenario, which is faster than the sublinear convergence rate $\mathcal{O} ( \frac{1}{\epsilon} ) $  achieved by ISTA-GS.

\subsection{Unfolding Structure III: ALPOM-GS}
\begin{figure}[tbp] 
	\centering
	\includegraphics[width=0.39\textwidth]{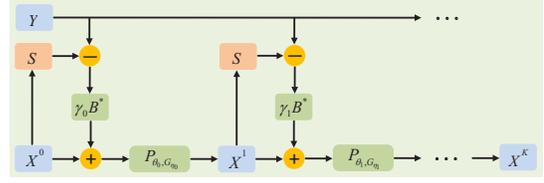}
	\captionsetup{font={footnotesize}}
	\caption{Illustration of proposed ALPOM-GS structure with trainable parameters $ \{ \gamma_k, \theta_k, \eta_k \}_{k=0}^{K-1}$.} 
	\label{ALPOM-GS}
	\vspace{-0.5cm}
\end{figure}

According to \cite{liu2019alista}, if weight matrix $\bm{B}^k$ can be obtained in advance, then fewer trainable parameters are required to be learned. 
Specifically, weight matrix $\bm{B}^k$ can be written as $ \bm{B}^k = \gamma_k \bm{B} $. 
We can solve the following optimization problem to obtain the weight matrix $ \bm{B}$ before the training stage
\begin{equation} \label{key16}
	\begin{aligned}
		& \mathop{\text{minimize}}_{\bm{B}\in \mathbb{R}^{2N\times2L}}  \quad \|   \bm{B}\tilde{\bm{S}}
		 \|_F^2 \\
		&\text{subject to} \quad \bm{B}_{i,:} \tilde{\bm{S}}_{:,i} = 1,\forall i \in [2N].
	\end{aligned}
\end{equation}
To obtain the optimal weight matrix, denoted as $ \bm{B}^*$, 
the projected gradient descend (PGD) method can be leveraged
to solve problem (\ref{key16}).
After obtaining $ \bm{B}^*$, we develop another unfolding neural network structure, termed ALPOM-GS, which can be written as follows
\begin{equation} \label{key17}
	\begin{aligned}
		\tilde{\bm{X}}^{k+1}\!=\!P_{\theta_k,G_{\eta_k}}\left(\tilde{\bm{X}}^k\! +\! \gamma_k \bm{B}^*(  \tilde{\bm{Y}}\! -\! \tilde{\bm{S}} \tilde{\bm{X}}^k ) \right)\!,\!  k\! =\! 0,\ldots,K-1,
	\end{aligned}
\end{equation}
where $ \bm{\Theta} = \{  \gamma_k, \theta_k, \eta_k \}_{k=0}^{K-1} $ are trainable parameters.
Note that we solve problem (\ref{key16}) for obtaining the optimal weight matrix $\bm{B}^*$ prior to the training stage. 
Fig. \ref{ALPOM-GS} shows the structure of the proposed ALPOM-GS. 
%For ALPOM-GS with $K$ layers, the total number of trainable parameters is $3K$. 
%As all the trainable parameters are scalars, the training process can be significantly simplified. 

\subsection{Training and Testing Strategies}
With supervised learning, we denote $ \{ \tilde{\bm{X}}_i^* , \tilde{\bm{Y}}_i \} _{i=1}^T $ as the set of training samples, where $ \tilde{\bm{X}}_i^*$ is label, $\tilde{\bm{Y}}_i$ is the data, and $T$ denotes the size of the training batch. 
The inputs of $K$-layer RNN include $ \tilde{\bm{Y}_i } $ and $ \tilde{\bm{X}}^0$.
The output of $K$-layer RNN can be expressed as $ \tilde{\bm{X}}^{K}(\bm{\Theta },\tilde{\bm{Y}}_i,\tilde{\bm{X}}^0) $.
Given $ \{ \tilde{\bm{X}}_i^* , \tilde{\bm{Y}}_i \} _{i=1}^T $, we solve the following problem to obtain the parameters of the $K$-layer RNN
\begin{equation} \label{key18}
\begin{aligned}
	\bm{\Theta}^* = \text{arg} \mathop{\text{min}}_{\bm{\Theta }} \sum_{i=1}^{T}\left\| \tilde{\bm{X}}^{K}(\bm{\Theta },\tilde{\bm{Y}}_i,\tilde{\bm{X}}^0)-\tilde{\bm{X}}^*_i  \right\|_F^2.
\end{aligned}
\end{equation}
\begin{figure*}[!htbp] 
	\setlength{\abovecaptionskip}{0.cm}
	\setlength{\belowcaptionskip}{0.cm}
	\centering 
	\captionsetup{font={footnotesize}}
	\begin{minipage}[t]{0.23\textwidth} 
		\centering 
		\includegraphics[width=4.6cm]{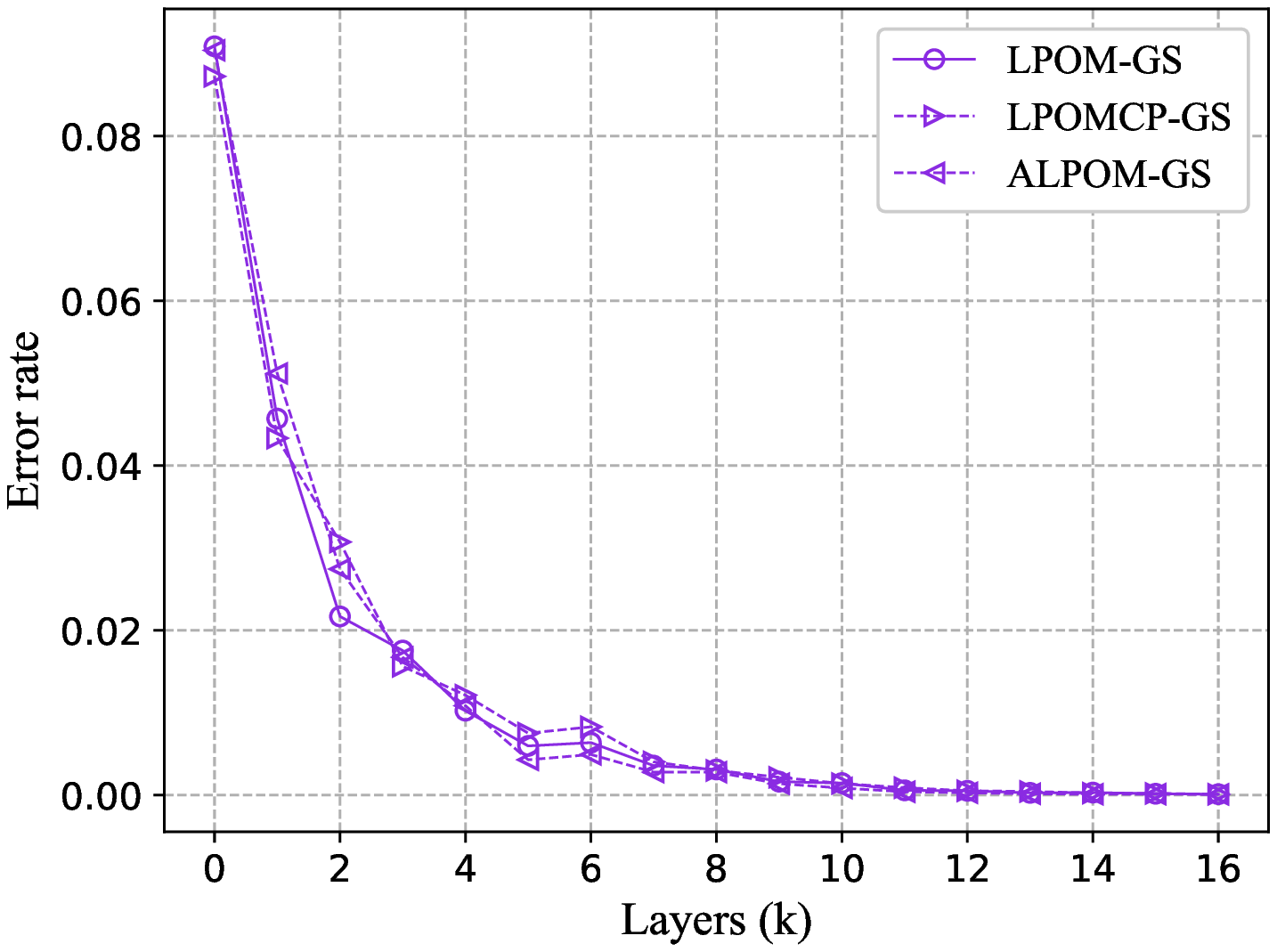} 
		\caption{Error rate of proposed methods.}
		\label{fig2}
	\end{minipage} \ 
	\begin{minipage}[t]{0.23\textwidth}  
		\centering 
		\includegraphics[width=4.6cm]{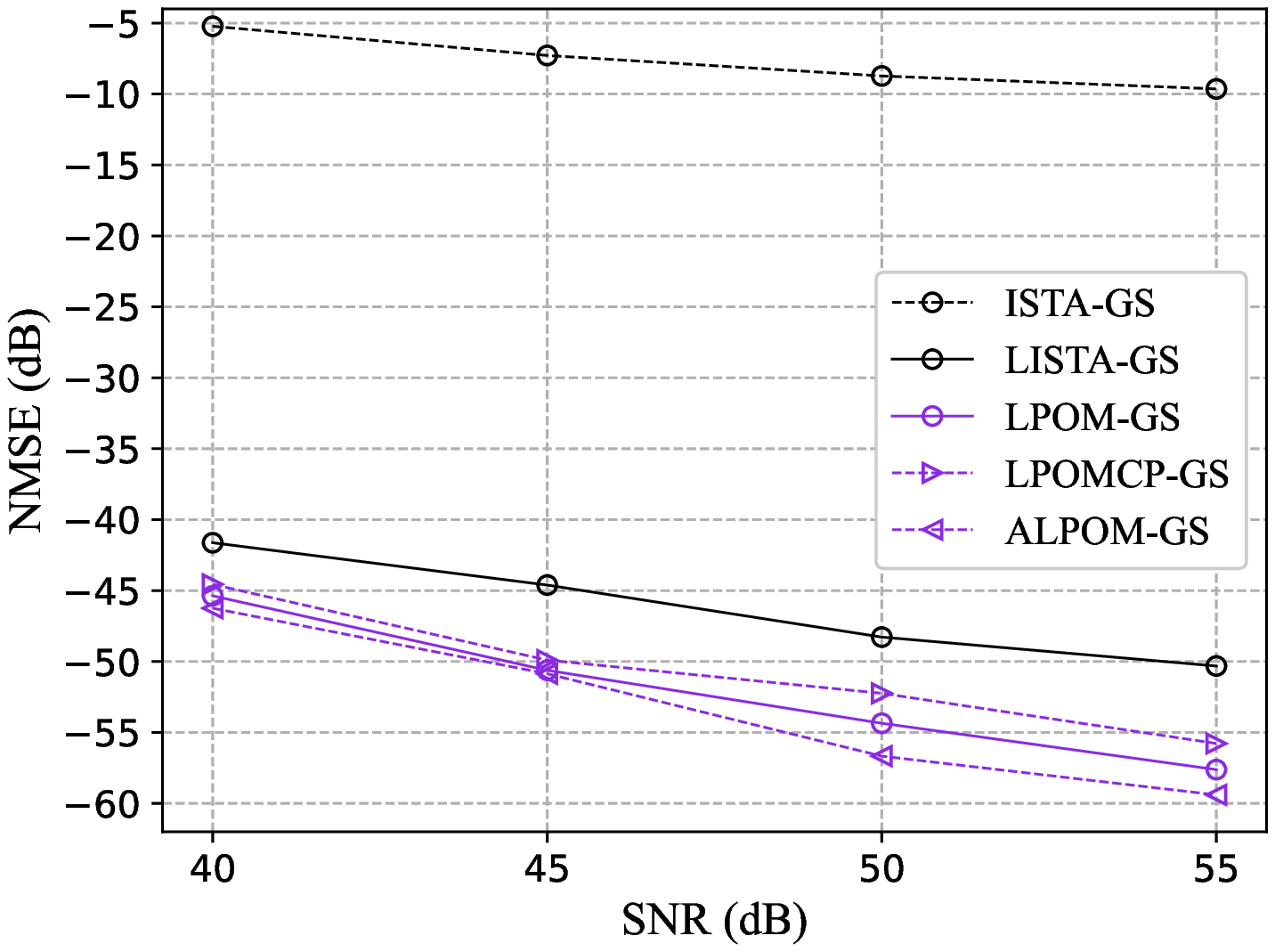} 
		\caption{NMSE versus SNR when  $\bm{S}$ is complex Gaussian matrix.}
		\label{fig3}
	\end{minipage}   \ 
	\begin{minipage}[t]{0.23\textwidth} 
		\centering 
		\includegraphics[width=4.6cm]{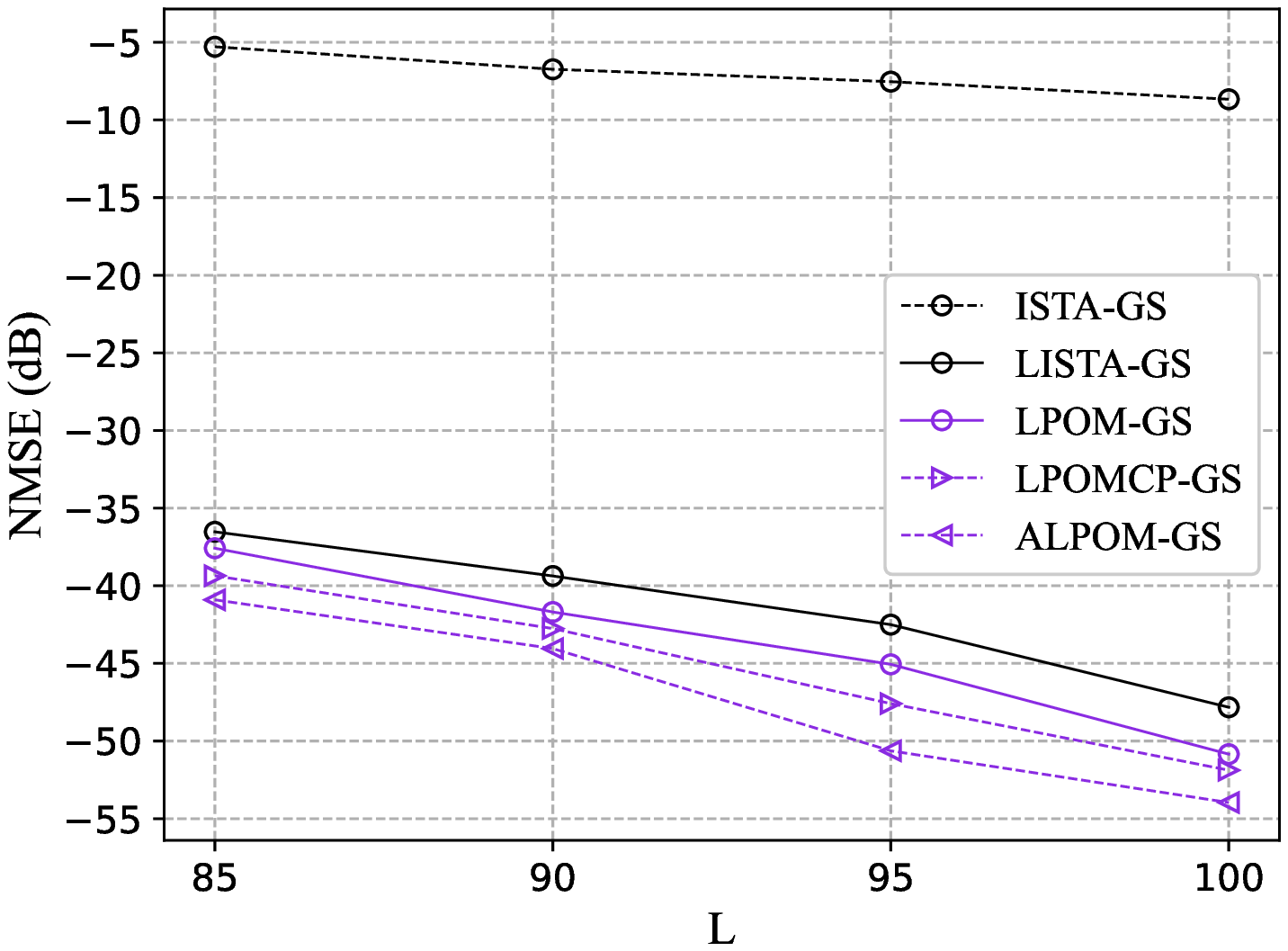} 
		\caption{NMSE versus L when   
			 $\bm{S}$ is binary matrix when SNR = 50 dB.}
		\label{fig4}
	\end{minipage} \ 
	\begin{minipage}[t]{0.23\textwidth}  
		\centering 
		\includegraphics[width=4.6cm]{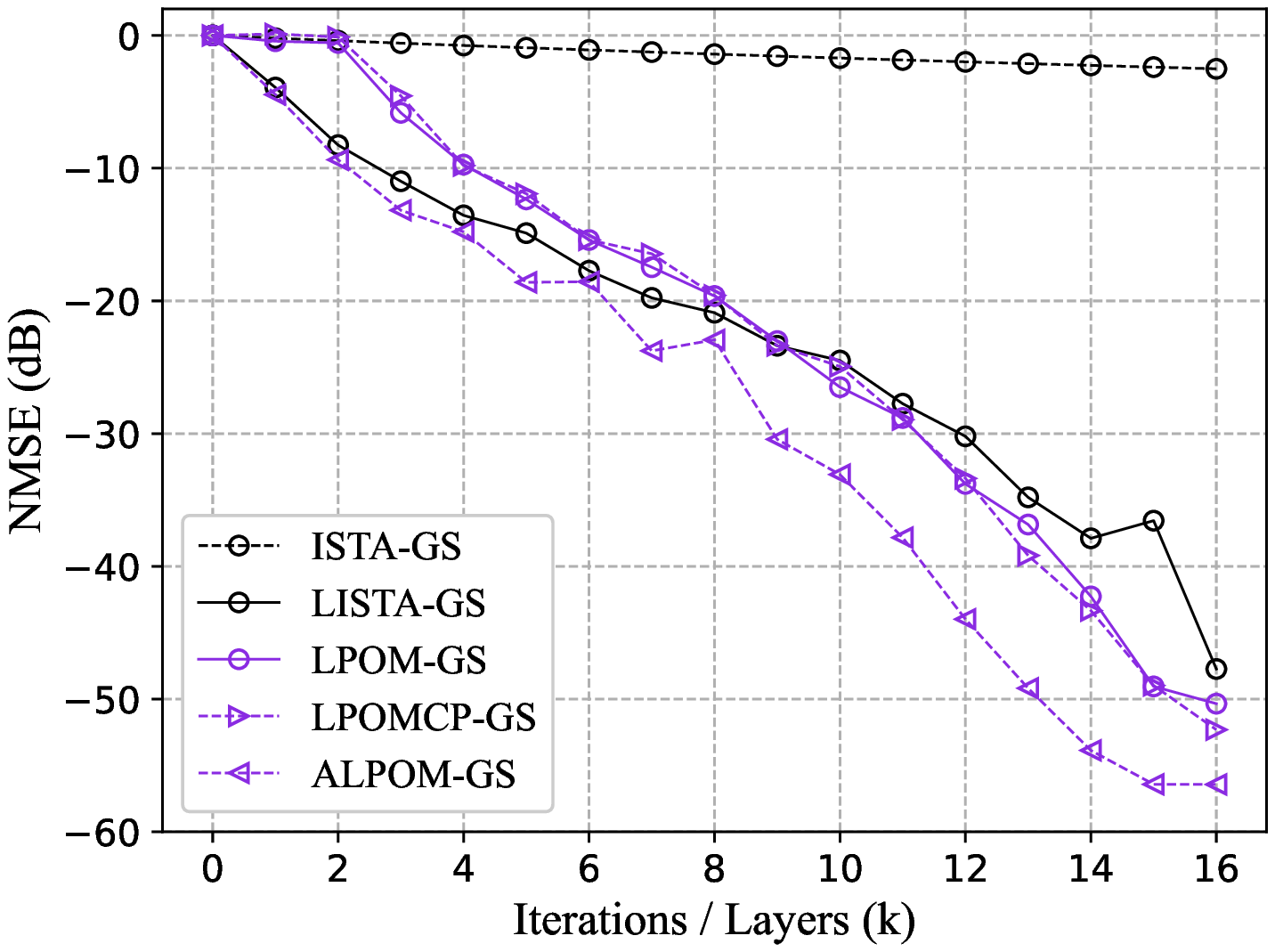}
		\caption{Performance comparison when $\kappa$ =5 and SNR = 50 dB. } 
		\label{fig5}
	\end{minipage} 
	\vspace{-0.1cm}
\end{figure*}
The weight matrices  $\bm{W}^k$ and $\bm{B}^k$ are initialized as $\bm{I}-\gamma_k\tilde{\bm{S}}^T\tilde{\bm{S}}$ and $\gamma_k\tilde{\bm{S}}^T$, respectively. 
After the initialization, we utilize the layer-by-layer training strategy to train the network parameters, so as to mitigate the probability of converging to the local minimum \cite{borgerding2017amp}. 
Specifically, after training the parameters of the first $k-1$ layers, denoted as $\bm{\Theta}_{0:k-2}$, we solve the unconstrained problem
$
\mathop{\text{min}}_{\bm{\Theta }_{k-1}}
\sum_{i=1}^{T}\left\| \tilde{\bm{X}}^{k}(\bm{\Theta }_{0:k-1},\tilde{\bm{Y}}_i,\tilde{\bm{X}}^0)-\tilde{\bm{X}}^*_i  \right\|_F^2,
$ to train the parameters of the $k$-th layer, denoted as $\bm{\Theta}_{k-1}$. 
Subsequently, we solve problem 
$
\mathop{\text{min}}_{\bm{\Theta }_{0:k-1}}
\sum_{i=1}^{T}\left\| \tilde{\bm{X}}^{k}(\bm{\Theta }_{0:k-1},\tilde{\bm{Y}}_i,\tilde{\bm{X}}^0)-\tilde{\bm{X}}^*_i  \right\|_F^2
$
to tune the parameters of the first $k$ layers (i.e., $\bm{\Theta}_{0:k-1}$).
Thus far, we finish the training of the $k$-layer neural network.

After all $K$ layers have completed the aforementioned process, we obtain all the trainable parameters. 
With the learned parameters, the proposed unfolding structures can be applied at the BS to perform JADCE after receiving the new signals in the testing stage. 

%Then in the validation and testing stage, given a new received signal $ \tilde{\bm{Y}}^{'} $, the learned unrolled network is applied for group-sparse matrix estimation by using $ \tilde{\bm{X}}^{k+1}=P_{\theta_k^*,F_{\eta_k^*}}\left((\bm{W}^k)^* \tilde{\bm{X}}^k + (\bm{B}^k)^*\tilde{\bm{Y}}^{'}  \right)  $, where$ (\bm{W}^k)^* $, $ (\bm{B}^k)^* $, $ \theta_k^* $ and $\eta_k^*$ are learned parameters in the $k$-th layer. 

\section{Simulation Results}
%In this section, we present the simulation results of the proposed three unfolding structures based on the proximal operator method for grant-free random access in IoT networks. 

%\subsection{Parameter Setting and Performance Metric}
In the simulations, $L$, $N$, and $M$ are set to be 100, 200, and 2, respectively. 
We consider independent Rayleigh fading channels and assume that the receiver noise follows the complex Gaussian distribution with zero mean and variance $\sigma^2$. 
The activity of each device follows an independent Bernoulli distribution, i.e., $\mathbb{P}(a_n=1)=0.1$ and $\mathbb{P}(a_n=0)=0.9$, $\forall\  n \in [N]$. 
According to (\ref{key3}), the transmit signal-to-noise ratio (SNR) is defined as $\frac{\mathbb{E}[\left\| \bm{SX} \right\|^2_F]}{\mathbb{E}[\left\| \bm{Z} \right\|^2_F]}
$. In the training stage, all neural networks have $K=16$ layers and the training data $ \{ \tilde{\bm{X}}_i^* , \tilde{\bm{Y}}_i = \tilde{\bm{S}}\tilde{\bm{X}^*_i}+\tilde{\bm{Z}} \} _{i=1}^{64} $ are given. In the testing stage, we generate a test set of $1000$ data samples and evaluate the group-sparse-matrix recovery performance in terms of the normalized mean square error  $
\text{NMSE}(\tilde{\bm{X}^k},\tilde{\bm{X}}^*)= 10\text{log}_{10}\left(\frac{\mathbb{E}\| \tilde{\bm{X}}^k-\tilde{\bm{X}}^* \|^2_F}{\mathbb{E}\| \tilde{\bm{X}}^* \|^2_F} \right)$.

\begin{figure*}[htbp] 
	\centering 
	\captionsetup{font={footnotesize}}
	%\begin{minipage}[t]{0.3\textwidth}  
	%	\centering 
	%	\includegraphics[width=4.7cm]{Gaussian_S50_con2.eps} 
	%	\caption{NMSE of the proposed and baseline methods when condition number $\kappa$ =2 and SNR = 50 dB. }
	%	\label{fig6}
	%\end{minipage}  
	\quad
	
	\quad
	%\begin{minipage}[t]{0.3\textwidth}  
	%	\centering
	%	\includegraphics[width=4.7cm]{Gaussian_S50_con25.eps} 
	%	\caption{NMSE of proposed and baseline methods when condition number $\kappa$ =25 and SNR = 50 dB. } 
	%	\label{fig8}
	%\end{minipage} 
\end{figure*}

With different preamble signature matrices, we compare the proposed three unfolding neural network structures with several baselines that address group-row-sparsity recovery, including ISTA-GS \cite{yuan2006model} and LISTA-GS \cite{shi2020sparse}.

Firstly, we train and test the proposed three network structures to validate  Theorem 1. 
The signature sequence matrix is generated according to the complex Gaussian distribution, i.e., $ \bm{S}\sim \mathcal{CN}(0,\bm{I})$, and we set SNR = 50 dB. 
The average error rate of each layer is numerically evaluated in terms of $\frac{1}{2VN} \sum_{i=1}^{V} \| \psi(\tilde{\bm{X}}^k_i) - \psi(\tilde{\bm{X}}^*_i) \|_1 $, where $V$ denotes the size of the test set. As shown in Fig. \ref{fig2}, our proposed structures can achieve accurate support recovery and device activity detection. 

Secondly, we evaluate the performance  of our proposed methods with that of the state-of-the-art methods. All the settings except SNR are the same as that for Fig. \ref{fig2}. 
Fig. \ref{fig3} shows the impact of SNR on NMSE. We observe that as SNR decreases, NMSE increases monotonically due to the increase of noise. By utilizing MCP that has a greater ability to induce sparsity, our proposed network structure improves the NMSE up to $18\%$ over LISTA-GS when SNR = 55 dB.

%Fig. \ref{fig2} shows the NMSEs of the proposed three unfolding structures and the baseline methods when the transmit SNR is $60$ dB. 
%It can be observed that, as the number of layers/iterations increases, the NMSE of all methods decreases monotonically until converging to certain values. 
%In addition, the proposed neural structures (i.e., LPOM-GS, LPOM-GSCP, and ALPOM-GS) converge with a faster rate and achieve a lower NMSE performance than the baseline methods (i.e., ISTA-GS, LISTA-GS, LISTA-GSCP, and ALISTA-GS). 
%This is because the MCP regularizer adopted in this paper has a greater ability in inducing sparsity than $\ell_1$-norm. 

%Fig. \ref{fig3} illustrates the impact of the transmit SNR on the recovery performance of the proposed unfolding neural networks for massive connectivity in IoT networks. 
%With the increase of the transmit SNR, the NMSE of all methods under consideration reduces, as the detrimental effect of noise is decreasing. 
%Moreover, the proposed unfolding structures with more trainable parameters significantly outperform the baseline methods in terms of the group-sparse-matrix estimation performance when the transit SNR is high. 

Thirdly, by randomly setting each entry of matrix $\bm S$ as $1$ or $-1$, we generate a binary signature sequence matrix $\bm{S}$. 
The results in Fig. \ref{fig4} indicate that our proposed network structures still obtain a lower NMSE under different lengths of the signature sequence, demonstrating the advantage in recovering the complex group-row-sparse signal in realistic wireless networks.  

%Figs. \ref{fig4} shows that the proposed unfolding structures achieve a better NMSE performance than ISTA-GS, LISTA-GS, and its variants when the signature sequence matrix $\bm{S}$ is a binary sequence matrix. These results demonstrate the robustness of our proposed unfolding neural networks as the superior performance can also be achieved in the case of non-Gaussian signature sequence matrix. 

Finally, the recovery performance of the proposed and baseline methods is compared when the signature sequence matrix $\bm{S}$ has a large condition number $\kappa$. 
We train all neural networks with ill-conditioned matrices $\bm{S}$ of condition number $\kappa = 5$. Fig. \ref{fig5} shows that although our proposed methods also suffer from ill-conditioning, they have a linear convergence rate and achieve better performance than other methods.
%To be specific, we generate matrix matrix $\bm{A} \in \mathbb{C}^{L \times N} \sim \mathcal{CN}(0,\bm{I})$, which is 
%decomposed as $ \bm{A} = \bm{U} \Sigma \bm{V}^* $ by singular value decomposition. 
%By replacing $\Sigma$ by a new $\Sigma^*$, we are able to obtain the signature sequence matrices of different %condition numbers. 
%Figs. \ref{fig7} shows the NMSE performance of the proposed and baseline methods when the condition numbers are  $\kappa = 15$.
%It can be observed that, with a larger condition number, the NMSE performance of all methods degrades due to the severer ill-conditioning of the signature sequence matrix. 
%However, the proposed unfolding structures always achieve a better NMSE performance than the baseline methods under different parameter settings. 

%First we sample a matrix $\bm{A} \in \mathbb{C}^{L \times N}$, i.e., $\bm{A} \sim \mathcal{CN}(0,\bm{I})$. Then, wedecompose $ \bm{A} = \bm{U} \Sigma \bm{V}^* $ by singular value decomposition and replace $\Sigma$ by a new $\Sigma^*$. So we can obtain the signature matrices of different condition number.

\section{Conclusions}
In this paper,  we investigated the JADCE problem in IoT networks with grant-free uplink transmission. 
The JADCE problem was formulated as a group-sparse-matrix estimation problem regularized by the non-convex MCP, which can be solved by using the proximal operator method. 
To reduce the computational complexity, we developed three unfolding neural network structures, which parameterize the algorithmic iterations. 
Simulations were conducted to evaluate the performance of the proposed method using different signature sequences. 
Results demonstrated that the proposed method can achieve better robustness, faster convergence rate, and higher estimation accuracy than the baseline methods.

\section*{Appendix}

\subsection{Proof of Theorem 1}
When $k=0$ and $\tilde{\bm{X}}^0 = \bm{0}$, we have $\text{supp}(\psi (\tilde{ \bm{X}}^{0})) = \emptyset \subseteq S $.
We assume $\tilde{\bm{X}}^k_{i,:}=\bm{0}$ for all $i \notin S $. According to (\ref{key12}), for $\forall j \in [M]$, we have
$
		\tilde{\bm{X}}^{k+1}_{i,j}  = \text{arg}\mathop{\text{min}}_{u} \quad \theta_{k}  g_{\eta_k} (u)  + \frac{1}{2} |u - v |^2  
$
where $ v = - \bm{B}^k_{i,:}\tilde{\bm{S}}_{:,l}(\tilde{\bm{X}}^k_{l,j} - \bm{X}^*_{l,j}) + \bm{B}^k_{i,:}\tilde{\bm{Z}}_{:,j}, \forall l\in S  $.
Since $ \theta_{k}  g_{\eta_k} (u)  + \frac{1}{2} |u - v |^2  \stackrel{(a)}{\geq}  \theta_k   \alpha |u| - \theta_k {\eta_k} u^2 + \frac{1}{2}|u-v|^2 = (\frac{1}{2}-\theta_k {\eta_k}) u^2 + \theta_k \alpha |x| - vu + \frac{1}{2}v^2 \stackrel{(b)}{\geq} (\frac{1}{2}-\theta_k {\eta_k}) u^2 + (\theta_k \alpha - |v|) |u| + \frac{1}{2}v^2 \stackrel{(c)}{\geq} \frac{1}{2}v^2 $, where (a) is based on\cite[Propety 1]{yang2019weakly}, (b) follows from the Cauchy–Schwarz inequality, and (c) holds according to (\ref{proof1}), we get $\tilde{\bm{X}}^{k+1}_{i,j}=0, \forall j \in [M] $, and thus $\tilde{\bm{X}}^{k+1}_{i,:}=\bm{0}$. By induction, we complete the proof. 

%\subsection{Proof of Lemma 1}
%
%
% We denote generalized gradient of $g_{\eta_k}(z)$ as $\partial g_{\eta_k}(z)$, then $ g_{\eta_k}^{'}(z) \leq \frac{g_{\eta_k}(z+\delta)-g_{\eta_k}(z)}{\delta} $ where $g_{\eta_k}^{'}(z) \in \partial g_{\eta_k}(z)$. Since $\frac{g_{\eta_k}(z)}{z}$ is non-increasing on $(0,+\infty)$ , we have $g_{\eta_k}^{'}(z) \leq \frac{g_{\eta_k}(z+\delta)-g_{\eta_k}(z)}{\delta}\leq \frac{g_{\eta_k}(z)}{z} \leq \alpha$. Besides, $-g_{\eta_k}^{'}(z) \leq \lim\limits_{\substack{\delta\rightarrow 0^+}} \frac{g_{\eta_k}(z-\delta)-g_\eta(z)}{\delta} \leq 0$ holds because  $g_{\eta_k}(z)$ is non-decreasing for $z \in (0,+\infty)$. In consequence, $0 \leq g_{\eta_k}^{'}(z) \leq \alpha, \forall z \in (0,+\infty)$. Similarly, for $z\in(-\infty,0)$, $ -\alpha \leq g_{\eta_k}^{'}(z) \leq 0$. Since $-\alpha \leq g_\eta^{'}(0) \leq \alpha$, then for any $z$, $g_{\eta_k}^{'}(z)^2 \leq \alpha^2$ holds. Finally $\| G^{'}_{\eta_k}(\bm{Z}) \|_F \leq \sqrt{\omega} \alpha$, where $\omega$ denotes the number of the elements of $\bm{Z}$.
%
\subsection{Proof of Theorem 2}
According to the structure of LPOMCP-GS and the optimality condition, for any $i \in S$, we have
\begin{equation}\label{proof2}
	\begin{aligned}
		\tilde{\bm{X}}^{k+1}_{i,:} - \tilde{\bm{X}}^{*}_{i,:} &= \tilde{\bm{X}}^{k}_{i,:} - \tilde{\bm{X}}^{*}_{i,:} - 
		\bm{B}^k_{i,:}\tilde{\bm{S}}_{:,S} 
		 (\tilde{\bm{X}}^k_{S,:}-\bm{X}^{*}_{S,:}) \\ &+  \bm{B}^k_{i,:}\tilde{\bm{Z}} - \theta_k G^{'}_{\eta_k}(\tilde{\bm{X}}^{k+1}_{i,:}).
	\end{aligned}
\end{equation}
where $G^{'}_{\eta_k}(\tilde{\bm{X}}^{k+1}_{i,:}) \in \partial G_{\eta_k}(\tilde{\bm{X}}^{k+1}_{i,:}) $ and $\partial G_{\eta_k}(\tilde{\bm{X}}^{k+1}_{i,:}) $ is the generalized gradient of  $ G_{\eta_k}(\tilde{\bm{X}}^{k+1}_{i,:}) $. 
Since $ \bm{B}_{i,:}\tilde{\bm{S}}_{:,i}=1$, (\ref{proof2}) can be expressed as 
\begin{equation}
	\begin{aligned}
		 \tilde{\bm{X}}^{k+1}_{i,:} - \tilde{\bm{X}}^{*}_{i,:} &= 
		-\mathop{\sum}_{j\in S , j\neq i } \bm{B}^k_{i,:}\tilde{\bm{S}} _{:,j}  (\tilde{\bm{X}}^k_{j,:}  -\bm{X}^{*}_{j,:}) \\ & 
		+  \bm{B}^k_{i,:}\tilde{\bm{Z}}  - \theta_k G^{'}_{\eta_k}(\tilde{\bm{X}}^{k+1}_{i,:}).
	\end{aligned}
	\label{proof28}
\end{equation}
For any $j \in [M]$, if $\tilde{\bm{X}}^{k+1}_{i,j},\delta >0$,  and $g_{\eta_k}^{'}(\tilde{\bm{X}}^{k+1}_{i,j}) \leq \frac{g_{\eta_k}(\tilde{\bm{X}}^{k+1}_{i,j}+\delta)-g_{\eta_k}(\tilde{\bm{X}}^{k+1}_{i,j})}{\delta} $ where $g_{\eta_k}^{'}(\tilde{\bm{X}}^{k+1}_{i,j}) \in \partial g_{\eta_k}(\tilde{\bm{X}}^{k+1}_{i,j})$ and $\partial g_{\eta_k}(\tilde{\bm{X}}^{k+1}_{i,j})$ is generalized gradient of $g_{\eta_k}(\tilde{\bm{X}}^{k+1}_{i,j})$. 
Since $\frac{g_{\eta_k}(\tilde{\bm{X}}^{k+1}_{i,j})}{\tilde{\bm{X}}^{k+1}_{i,j}}$ is non-increasing on $(0,+\infty)$ , we have $g_{\eta_k}^{'}(\tilde{\bm{X}}^{k+1}_{i,j}) \leq \frac{g_{\eta_k}(\tilde{\bm{X}}^{k+1}_{i,j}+\delta)-g_{\eta_k}(\tilde{\bm{X}}^{k+1}_{i,j})}{\delta}\leq \frac{g_{\eta_k}(\tilde{\bm{X}}^{k+1}_{i,j})}{\tilde{\bm{X}}^{k+1}_{i,j}} \leq \alpha$. Besides, $-g_{\eta_k}^{'}(\tilde{\bm{X}}^{k+1}_{i,j}) \leq \lim\limits_{\substack{\delta\rightarrow 0^+}} \frac{g_{\eta_k}(\tilde{\bm{X}}^{k+1}_{i,j}-\delta)-g_\eta(\tilde{\bm{X}}^{k+1}_{i,j})}{\delta} \leq 0$ holds because  $g_{\eta_k}(\tilde{\bm{X}}^{k+1}_{i,j})$ is non-decreasing for $\tilde{\bm{X}}^{k+1}_{i,j} \in (0,+\infty)$. In consequence, $0 \leq g_{\eta_k}^{'}(\tilde{\bm{X}}^{k+1}_{i,j}) \leq \alpha, \forall \tilde{\bm{X}}^{k+1}_{i,j} \in (0,+\infty)$. Similarly, if  $\tilde{\bm{X}}^{k+1}_{i,j} <0$, then $ -\alpha \leq g_{\eta_k}^{'}(\tilde{\bm{X}}^{k+1}_{i,j}) \leq 0$. Since $-\alpha \leq g_\eta^{'}(0) \leq \alpha$, then for any $\tilde{\bm{X}}^{k+1}_{i,j}$, $g_{\eta_k}^{'}(\tilde{\bm{X}}^{k+1}_{i,j})^2 \leq \alpha^2$ holds. Finally, we have $ \| G^{'}_{\eta_k}(\tilde{\bm{X}}^{k+1}_{i,:})\|_2 \leq \sqrt{M} \alpha$.

By taking norm on both sides of (\ref{proof28}), we get
\begin{equation}
	\begin{aligned}
		& \| \tilde{\bm{X}}^{k+1}_{i,:} - \tilde{\bm{X}}^{*}_{i,:} \|_2 \leq 
		\mathop{\sum}_{j\in S , j\neq i } | \bm{B}^k_{i,:}\tilde{\bm{S}} _{:,j} |  \| \tilde{\bm{X}}^k_{j,:}  -\bm{X}^{*}_{j,:} \|_2 
		\\ & + \| \bm{B}^k_{i,:}\tilde{\bm{Z}} \|_2 + \theta_k \| G^{'}_{\eta_k}(\tilde{\bm{X}}^{k+1}_{i,:})\|_2
		 \\& \leq \phi \mathop{\sum}_{j\in S , j\neq i } \| \tilde{\bm{X}}^k_{j,:}-\bm{X}^{*}_{j,:} \|_2 + \| \bm{B}^k_{i,:}\|_2 \| \tilde{\bm{Z}} \|_F   + \theta_k \sqrt{M} \alpha
	\end{aligned}
\end{equation}
According to Theorem 1, we have $\| \tilde{\bm{X}}^{k} - \tilde{\bm{X}}^{*} \|_{2,1} = \| \tilde{\bm{X}}^{k}_{S,:} - \tilde{\bm{X}}^{*}_{S,:} \|_{2,1}$, and
\begin{equation}
	\begin{aligned}
		&\| \tilde{\bm{X}}^{k+1} - \tilde{\bm{X}}^{*} \|_{2,1} = \mathop{\sum}_{i\in S} 	\| \tilde{\bm{X}}^{k+1}_{i,:} - \tilde{\bm{X}}^{*}_{i,:} \|_2  \\ &\leq \phi(|S|-1) \| \tilde{\bm{X}}^{k} - \tilde{\bm{X}}^{*} \|_{2,1} + |S|\theta_{k} \sqrt{M} \alpha + \epsilon \mu_{B}.
	\end{aligned}
\end{equation}
By taking supremum over $(\tilde{\bm{X}^*},\tilde{\bm{Z}})\in\mathcal{X}(\mu_x,s,\epsilon)$ on both sides, we have
\begin{equation}
	\begin{aligned}
		\mathop{\sup}_{(\tilde{\bm{X}^*},\tilde{\bm{Z}})} 	\| \tilde{\bm{X}}^{k+1} - \tilde{\bm{X}}^{*} \|_{2,1}  &\leq \phi(s-1) \mathop{\sup}_{(\tilde{\bm{X}^*},\tilde{\bm{Z}})} \| \tilde{\bm{X}}^{k} - \tilde{\bm{X}}^{*} \|_{2,1}  \\ & + s\theta_{k} \sqrt{M} \alpha + \epsilon \mu_{B}.
	\end{aligned}
\end{equation}
Since $\theta_k = 	\frac{1}{\alpha} \left(  \phi \mathop{\text{sup}}_{ (\tilde{\bm{X}^*},\tilde{\bm{Z}})  }  \| \tilde{\bm{X}}^k  - \tilde{ \bm{X}}^* \|_{2,1} + \mu_{B} \epsilon \right) $, if $\phi s \sqrt{M} + \phi s - \phi < 1$, then
\begin{equation}
	\begin{aligned}
		\mathop{\sup}_{(\tilde{\bm{X}^*},\tilde{\bm{Z}})} 	\| \tilde{\bm{X}}^{k+1} - \tilde{\bm{X}}^{*} \|_{2,1}  &\leq (\phi s \sqrt{M} + \phi s - \phi)^{k+1} s \mu_x   \\&+ \frac{\mu_{B}(s\sqrt{M}+1)\epsilon}{1+\phi-\phi s-\phi s \sqrt{M} }.
	\end{aligned}
\end{equation}
Let $ c_1 = -\text{log}(\phi s \sqrt{M} + \phi s - \phi) $ and $c_2 = \frac{\mu_{B}(s\sqrt{M}+1)\epsilon}{1+\phi-\phi s-\phi s \sqrt{M} } $. Since $\|\bm{\bm{X}}\|_F\leq\|\bm{X}\|_{2,1}$, we have
\begin{equation}
	\begin{aligned}
		\mathop{\sup}_{(\tilde{\bm{X}^*},\tilde{\bm{Z}})} 	\| \tilde{\bm{X}}^{k+1} - \tilde{\bm{X}}^{*} \|_F \leq s \mu_x \text{exp}(-c_1(k+1)) +\epsilon c_2. 
	\end{aligned}
\end{equation}

%In this paper, in order to solve JADCE problem in grant-free massive access for IoT networks and improve performance of neural networks for sparse reconstruction, we propose a new neural network named LPOM-GS, which is based on unrolled proximal operator algorithm. Compared with other regularizers, we choose MCP to have better approximation of the $\ell_0$ norm to induce sparsity. The three structures of proposed network have different number of trainable parameters. Furthermore, we consider the JADCE in grant-gree massive connectivity forr mMTC. Simulations demonstrate that LPOM-GS outperforms other group sparse recovery algorithms.
 
\bibliographystyle{IEEEtran}
\bibliography{ref}

\end{document}